\documentclass[12pt,a4paper]{article}
\usepackage[utf8]{inputenc}
\usepackage{amsmath}
\usepackage{placeins}
\usepackage{url}
\usepackage{svg}
\usepackage{natbib}
\usepackage{titlesec}
\usepackage{color,xcolor,setspace,geometry}
\setcitestyle{authoryear,open={(},close={)}}
\usepackage{cite}
\usepackage{soul}
\usepackage{bbm}
\usepackage{dsfont}
\soulregister\citep7
\soulregister\cite7
\soulregister\citet7
\soulregister\citealp7
\soulregister\ref7
\usepackage{amsfonts}
\usepackage{tikz}
\usetikzlibrary{shapes}
\usetikzlibrary{plotmarks}
\usetikzlibrary{tikzmark,matrix,calc}
\usetikzlibrary{arrows.meta,calc,decorations.markings,math,arrows.meta}
\usepackage{amssymb}
\usepackage{multirow}
\usepackage{todonotes}
\usepackage{graphicx}
\usepackage{array}
\usepackage{arydshln}
\usepackage[final]{changes}
\titleformat*{\section}{\large\bfseries}
\titleformat*{\subsection}{\bfseries}

\providecommand{\keywords}[1]
{
  \small	
  \textbf{\textit{Keywords---}} #1
}

\usepackage{natbib}
\usepackage{datatool}
\usepackage{etoolbox}

\DTLnewdb{bibnotes}

\def\bibnote#1#2{
  \DTLnewrow{bibnotes}
  \DTLnewdbentry{bibnotes}{mylabel}{#1}
  \DTLnewdbentry{bibnotes}{mynote}{#2}
}

\makeatletter
\patchcmd{\@lbibitem}
  {\item[\hfil\NAT@anchor{#2}{\NAT@num}]}
  {
    \item[\hfil\NAT@anchor{#2}{\NAT@num}]
    \DTLforeach[\DTLiseq{\mylabel}{#2}]{bibnotes}{\mylabel=mylabel,\mynote=mynote}{\text{\mynote}}
  }{}{\message{^^JPatching failed^^J}}

\makeatother
\begin{document}
\begin{spacing}{1.35}

\listofchanges  
\title{Nonparametric estimation of multivariate hidden Markov models using tensor-product B-splines}
\author{Rouven Michels\thanks{corresponding author: r.michels@uni-bielefeld.de} \thanks{ Bielefeld University, Universitätsstraße 25, 33615 Bielefeld, Germany}, Roland Langrock\footnotemark[2]}
\date{}

\maketitle

\begin{abstract} 
\noindent
For multivariate time series driven by underlying states, hidden Markov models (HMMs) constitute a powerful framework which can be flexibly tailored to the situation at hand. However, in practice it can be challenging to choose an adequate emission distribution for multivariate observation vectors. For example, the marginal data distribution may not immediately reveal the within-state distributional form, and also the different data streams may operate on different supports, rendering the common approach of using a multivariate normal distribution inadequate.   
Here we explore a nonparametric estimation of the emission distributions within a multivariate HMM based on tensor-product B-splines. In two simulation studies, we show the feasibility of our modelling approach and demonstrate potential pitfalls of inappropriate choices of parametric distributions. To illustrate the practical applicability, we present a case study where we use an HMM to model the bivariate time series comprising the lengths and angles of goalkeeper passes during the UEFA EURO 2020, investigating the effect of match dynamics on the teams' tactics.
\end{abstract}

\section{Introduction}

The versatility of hidden Markov models (HMMs) makes them suitable for analysing various types of time series. 
In particular, it is conceptually straightforward to extend the basic model formulation --- within which at each time point the state of an unobserved Markov chain selects which of finitely many univariate distributions generates the observation at that time (\citealp{visser2002fitting,zucchini}) --- to address multivariate time series, e.g.\ bivariate time series of step lengths and turning angles in animal movement modelling \citep{beumer2020application} or high-dimensional financial time series in portfolio management \citep{fiecas2017shrinkage}. In terms of the model formulation, this involves choosing an appropriate {multivariate} emission distribution for the sequence of observation {vectors}.  

The simplest and most common approach to deal with multivariate observation vectors within HMMs is to assume contemporaneous conditional independence (see, e.g.\ \citealp{altman2007mixed, deruiter2017multivariate, van2019classifying}), i.e.\ that the elements of the observation vector are conditionally independent of each other, given the states. In that case, the multivariate emission distribution reduces to a simple product of univariate distributions, such that for each dimension a suitable univariate parametric family can be chosen. The obvious caveat of this approach is that potential within-state dependence of the different data streams is neglected, which can be highly problematic for example when using HMMs for forecasting in finance (as concentration risks would not be adequately captured). 
The standard way to avoid this strong assumption is to use the multivariate normal (see, e.g.\ \citealp{ailliot2009space, spezia2010bayesian,phillips2015objective, punzo2016clustering}) or --- very rarely --- other multivariate parametric distributions (see, e.g.\ \citealp{bulla2012multivariate,orfanogiannaki2018multivariate,ngo2019understanding}). 
For settings in which it cannot reasonably be assumed that the marginal distributions of the different data streams come from the same distributional family, copulas can be used to flexibly model the within-state dependence structure by stitching together univariate marginal distributions, possibly from different distributional families (\citealp{brunel2005unsupervised, lanchantin2011unsupervised, hardle2015hidden, otting2021copula, zimmerman2022copula}). 

The latter two approaches offer different ways to capture within-state dependence and will in most applications suffice, however they share the caveat that assumptions on the distributional form need to be made (for the copula-based approach, the copula also needs to be chosen). Prior to modelling, there is however no way to conduct an exploratory data analysis \textit{within} states to explore their empirical distribution, rendering it challenging to select an adequate distributional family. These difficulties exist also in the univariate case (see, e.g.\ \citealp{langrock2015nonparametric, langrock2018spline}), but they are exacerbated in the multivariate setting due to the additional challenge of how to model the dependence between the different data streams. To avoid the potential pitfalls associated with an unfortunate choice of the multivariate emission distribution --- potentially poor fit, invalid inference on the number of states, and imprecise state decoding, to name but a few --- we here explore an alternative nonparametric approach. Specifically, we discuss using multivariate tensor-product B-splines to estimate multivariate emission distributions in a data-driven way, i.e.\ without the need to make any distributional assumptions. This approach can relatively easily be implemented and used to nonparametrically fit HMMs to 2-- or 3--dimensional time series, while for higher dimensions it will typically not be feasible due to the curse of dimensionality.

In two simulation studies, we demonstrate the feasibility of the suggested approach for low-dimensional multivariate time series, and discuss in which type of scenarios it may be worth to adopt such a nonparametric technique.  
We further illustrate our approach in a real-data case study, modelling bivariate data on the length and angle of goalkeeper passes during the UEFA European Football Championship 2020 (played in 2021). We find teams to switch between different strategies in the build-up, and illustrate how the incorporation of covariates into the state-switching probabilities can reveal potential tactical adjustments by the team managers.

\section{Multivariate hidden Markov models}

\subsection{Model formulation and dependence assumptions}

An HMM comprises an unobserved first-order Markov chain $\{ g_t\}_{t=1}^T$, with $g_t \in \{1,\ldots,N\}$, and a $D$-dimensional observable time series $\{\boldsymbol{y}_t\}_{t=1}^T$, with $\boldsymbol{y}_t = (y_{t1}, \ldots, y_{tD})$. 
The Markov chain evolves according to the initial distribution $\boldsymbol{\delta}=\bigl( \Pr(g_1=1), \ldots, \Pr(g_1=N) \bigr)$ and the transition probability matrix (t.p.m.) $\boldsymbol{\Gamma} = (\gamma_{ij}),$ with $\gamma_{ij} = \Pr(g_t = j| g_{t-1} = i), \ i,j = 1, \ldots, N$. The distribution of the observed dependent variable $\boldsymbol{y}_t$ is fully determined by the underlying state at time $t$, i.e.\ it is assumed that 
\begin{equation*}
f(\boldsymbol{y}_t|g_1, \ldots, g_T, \boldsymbol{y}_1, \ldots, \boldsymbol{y}_{t-1},\boldsymbol{y}_{t+1},\ldots,\boldsymbol{y}_T) = f(\boldsymbol{y}_t|g_t),
\end{equation*}
with $f$ either a density or a probability mass function depending on the type of data to be modelled. In other words, the observation $\boldsymbol{y}_t$ is assumed to be conditionally independent of former states and observations, given $g_t$. In the existing literature, the multivariate emission distributions $f(\boldsymbol{y}_t|g_t)$, for $g_t = 1,\ldots,N$, a) are assumed to be the products of univariate distributions (when assuming contemporaneous conditional independence), b) are taken from a parametric family of multivariate distributions (typically the multivariate normal), or c) are built using copulas. In the next section, we propose a fourth option, namely the nonparametric estimation of $f(\boldsymbol{y}_t|g_t)$, which is applicable only in case the observations in each of the $D$ dimensions are continuous-valued.

\subsection{Spline-based construction of the multivariate distributions}

We propose to use tensor-product B-splines to effectively nonparametrically construct the multivariate emission distributions within an HMM. Tensor-product B-splines are simply products of univariate B-splines, the latter being constructed by connecting low-order polynomials, typically quadratic or cubic, at pre-defined knots \citep{splines,eilers2006fast}. 
Specifically, following \citet{epub31413}, we construct the emission distribution as
\begin{equation}\label{eq1}
    f(\boldsymbol{y}_t |g_t = i) = \sum_{j_1 = 1}^{n_1} \ldots \sum_{j_D = 1}^{n_D} a_{j_1, \ldots, j_D, i} B_{j_1}^{(1)} (y_{t1}) \cdots B_{j_D}^{(D)} (y_{tD}), \ \ \ i = 1, \ldots, N,
\end{equation}
where $n_d,\ d = 1, \ldots, D$, denotes the number of basis functions in each dimension (equally spaced in the respective support) 
and $a_{j_1, \ldots, j_D, i},\ j_d = 1, \ldots, n_d
$ are the coefficients to be estimated (a set of $\prod_{i=1}^{D} n_i$ coefficients for each of the $N$ states). In particular, $a_{j_1, \ldots, j_D, i},\ j_d = 1, \ldots, n_d,
$ has a $D$-dimensional structure, e.g.\ a matrix structure in two dimensions.  
While this model formulation still involves a finite-dimensional parameter space, the use of a relatively large number of basis functions in each dimension implies effectively unlimited flexibility. Moreover, the estimated coefficients $a_{j_1, \ldots, j_D, i}$ are neither of interest nor interpretable on their own, such that B-spline-based inference is typically classified as a nonparametric approach (see, e.g., \citealp{ruppert2009semiparametric}).
We would typically use cubic polynomials as basis functions, since they are twice continuously differentiable and hence yield smooth estimates \citep{splines}.

For the construction as in (\ref{eq1}) to meet the requirements of a multivariate distribution function, we scale the B-spline basis functions $B_{\cdot}^{(\cdot)}$ such that each of them integrates to one, and constrain the coefficients $a_{j_1, \ldots, j_D, i}$ 
such that $\sum_{j_1 = 1}^{n_1} \ldots \sum_{j_D = 1}^{n_D} a_{j_1, \ldots, j_D, i} = 1$ for $i=1,\ldots,N$, and $a_{j_1, \ldots, j_D, i} \geq 0$ for $j_d = 1, \ldots, n_d,\ d = 1, \ldots, D,\ i = 1, \ldots, N$.
To facilitate meeting the latter two constraints, we consider a re-parameterisation using the multinomial logit link, 
\begin{equation*}\label{logitlink}
a_{j_1, \ldots, j_D, i} = \frac{\exp(\beta_{j_1, \ldots, j_D, i})}{\sum_{k_1 = 1}^{n_1} \ldots \sum_{k_D = 1}^{n_d} \exp(\beta_{k_1, \ldots, k_D, i})},
\end{equation*}
for $j_d = 1, \ldots, n_d,\ d = 1, \ldots, D,\ i = 1, \ldots, N$. We then estimate the unconstrained parameters $\beta_{j_1, \ldots, j_D, i}$, fixing one of the $D$ coefficients to 0 (reference category).

\subsection{Maximum likelihood estimation}

The model parameters can relatively straightforwardly be estimated using numerical likelihood maximisation. To calculate the likelihood function, we use the forward algorithm (\citealp{zucchini}), which is associated with a computational cost that is (only) linear in $T$ due to applying recursive computing. We first define the forward variables as
\begin{equation*}
    \boldsymbol{\alpha}_t = \bigl(\alpha_t(1), \dots, \alpha_t(N) \bigr), \ \text{with} \ \alpha_t(i) = f(\boldsymbol{y}_1,\ldots,\boldsymbol{y}_t, g_t = i), \ i = 1, \ldots, N, \ t = 1, \ldots, T.
\end{equation*}
By construction, $\boldsymbol{\alpha}_t$ contains information on the likelihood of the data up to time $t$ and additionally on the state probabilities at time $t$. The forward variables can be updated recursively, 
$$\boldsymbol{\alpha_1} = \delta \mathbf{P}_1, \quad \boldsymbol{\alpha_t} = \boldsymbol{\alpha_{t-1}} \boldsymbol{\Gamma} \mathbf{P}_t, \quad t=2,\ldots,T,$$
where $\mathbf{P}_t = $ diag$\bigl(f(\boldsymbol{y}_t| g_t = 1),\ldots,f(\boldsymbol{y}_t| g_t = N)\bigr)$.
The likelihood for the full HMM sequence of a single time series is then obtained as 
\begin{equation}\label{likelihood}
\mathcal{L} = \boldsymbol{\alpha_T}\mathbf{1}' = \delta \mathbf{P}_1 \boldsymbol{\Gamma} \mathbf{P}_2 \cdot \ldots \cdot \boldsymbol{\Gamma} \mathbf{P}_T \mathbf{1}',
\end{equation}
where $\mathbf{1} \in \mathbb{R}^N$ is an $N$-dimensional row vector of ones. In the following, we assume $\delta$ to be the stationary distribution of the Markov chain, such that rather than estimating the initial distribution, it is taken as the solution to $\delta \boldsymbol{\Gamma} = \delta$ subject to $\sum_{i=1}^N \delta_i = 1$ and $\delta_i \geq 0$, $i = 1, \ldots, N$. 
 
When numerically optimising the likelihood in (\ref{likelihood}), a key challenge in practice is the risk to miss the global maximum. The common strategy of trying many different starting values in the optimisation is difficult to implement in our setting, as the specification of the coefficients $a_{j_1, \ldots, j_D, i}$ is not intuitive. Alternatively, the initial values can be chosen more systematically, running $k$-means clustering (see, e.g., \citealp{epub31413}), with $k=N$, on the multivariate observation vectors, then separately fitting $D$-dimensional distributions to the resulting clusters (using the same spline formulation as within the HMM). The resulting estimates can serve as initial values for the optimisation of the HMM likelihood.

\subsection{Addressing the bias-variance trade-off}

To arrive at a good balance between underfitting and overfitting, we could use very large numbers of basis functions but then penalise wiggliness \citep{eilers2006fast}, or alternatively simply choose a moderate number of basis functions in each dimension, without wiggliness penalty in the objective function. 
In the multivariate setting, both of these approaches suffer from the curse of dimensionality --- either the smoothing parameters or the numbers of basis functions need to be chosen from a $D$--dimensional grid. However, with the penalisation approach there is a second curse of dimensionality in that the number of coefficients to be estimated also increases exponentially in $D$, which given the numerical likelihood optimisation is anything but ideal.  
Thus, in our simulations and the case study below, we implemented the approach without penalisation, choosing appropriate numbers of basis functions using cross-validation. 

\subsection{Cross-validation for time series data in HMMs}

\citet{celeux2008selecting} consider two different ways to implement cross-validation for a single time series: two-fold half-sampling, where the observations with even (and then also those with odd) indices in the original time series are used for training and the remaining observations for testing, with the advantage that the Markov property is retained for the subsampled data,
or alternatively multi-fold random selection of the training data, then treating the omitted observations as missing. In our simulations below, we implement the latter. 
In applications involving multiple time series, e.g.\ the one considered in Section \ref{casestudy},
the splitting into training and test data can most conveniently be performed at the level of the different sequences, i.e.\ using some of the time series for training and the others for testing. 

\section{Simulation study}\label{chap:models}
In the following, we demonstrate the feasibility of the nonparametric estimation of multivariate HMMs in two simulation studies, and explore and showcase situations in which either approach --- parametric or nonparametric --- can be preferable. In the first simulation experiment, we consider a situation where standard parametric approaches are not able to capture a given complex distributional shape, such that the nonparametric approach will lead, {\it inter alia}, to improved state decoding. In the second experiment, we then explore the relative performance of the nonparametric approach in a setting where the parametric model  coincides with the data-generating process, such that the additional flexibility provided by the nonparametric approach is in fact no needed. This is done in order to provide an idea of the possible costs associated with applying the more flexible approach.

\subsection{Bivariate correlated gamma emission distribution}

\begin{figure}[!htb]
\centering
\includegraphics[width= 0.9\textwidth]{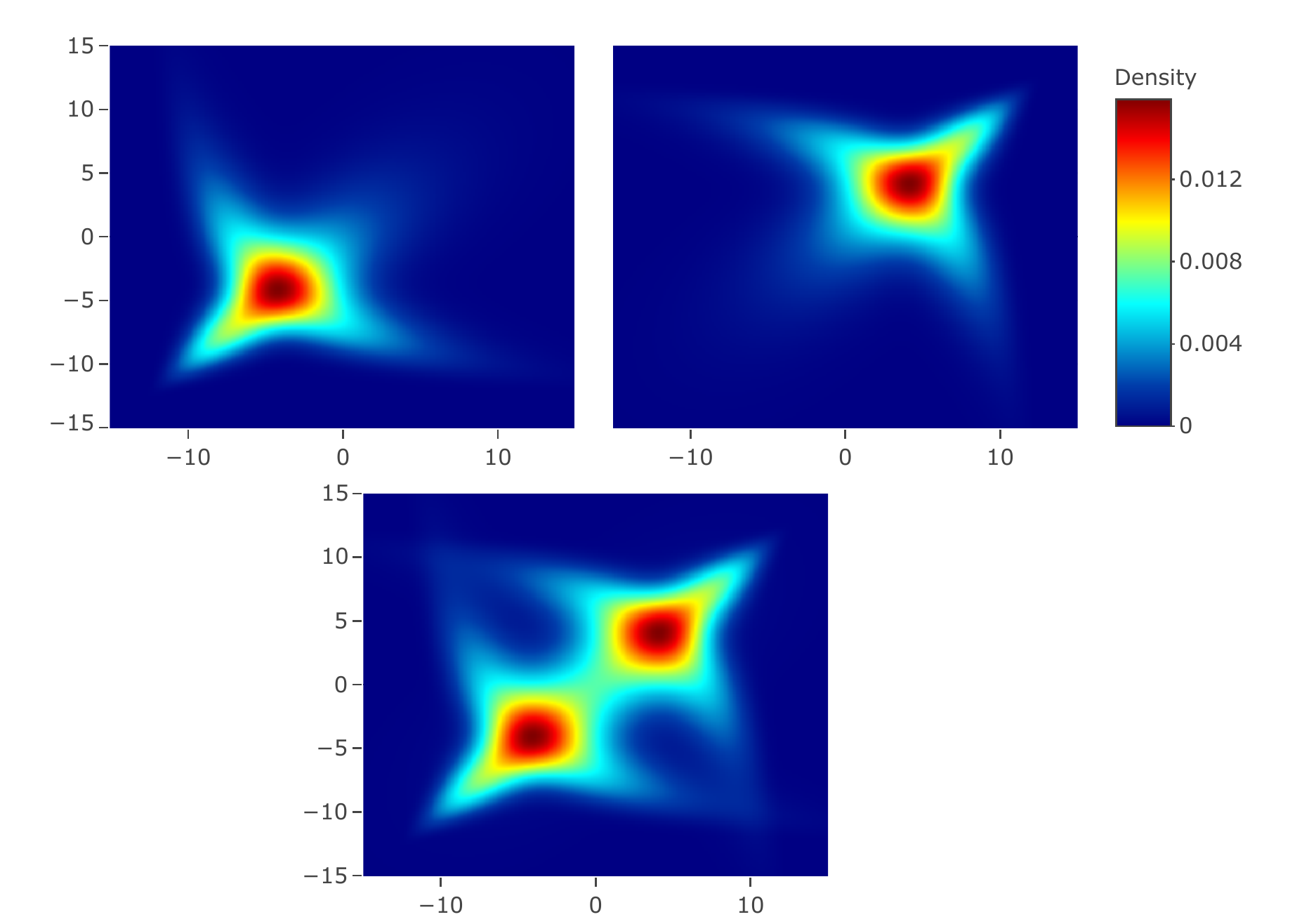}
\caption{\label{simdata} Contour plots of the true emission distributions (top~left~$\hat{=}$~state 1, top right $\hat{=}$ state 2) and the corresponding mixture distribution (bottom plot).}
\end{figure}

We first consider a setting with relatively complex bivariate emission distributions. 
The states of the Markov chain were generated using a $2 \times 2$ t.p.m.\ with both diagonal entries equal to $0.97$, resulting in high persistence in both states. Conditional on either state, we simulated the bivariate observations from a distribution
constructed based on two gamma-distributed marginals, stitched together using a Gaussian copula (Figure~\ref{simdata}).
Due to the overlap of the two emission distributions, neither the shape nor the dependence structure within states will be evident based on exploratory data analysis. Indeed, a standard bivariate Gaussian HMM might be considered to be an adequate model for such data. Using 100 simulations runs with $T = 2000$ observations per run, we explore the consequences of indeed using a bivariate Gaussian HMM for such data, and compare
the results to those obtained using the more flexible nonparametric approach.

\begin{figure}[!h]
\centering
\includegraphics[scale = 0.4]{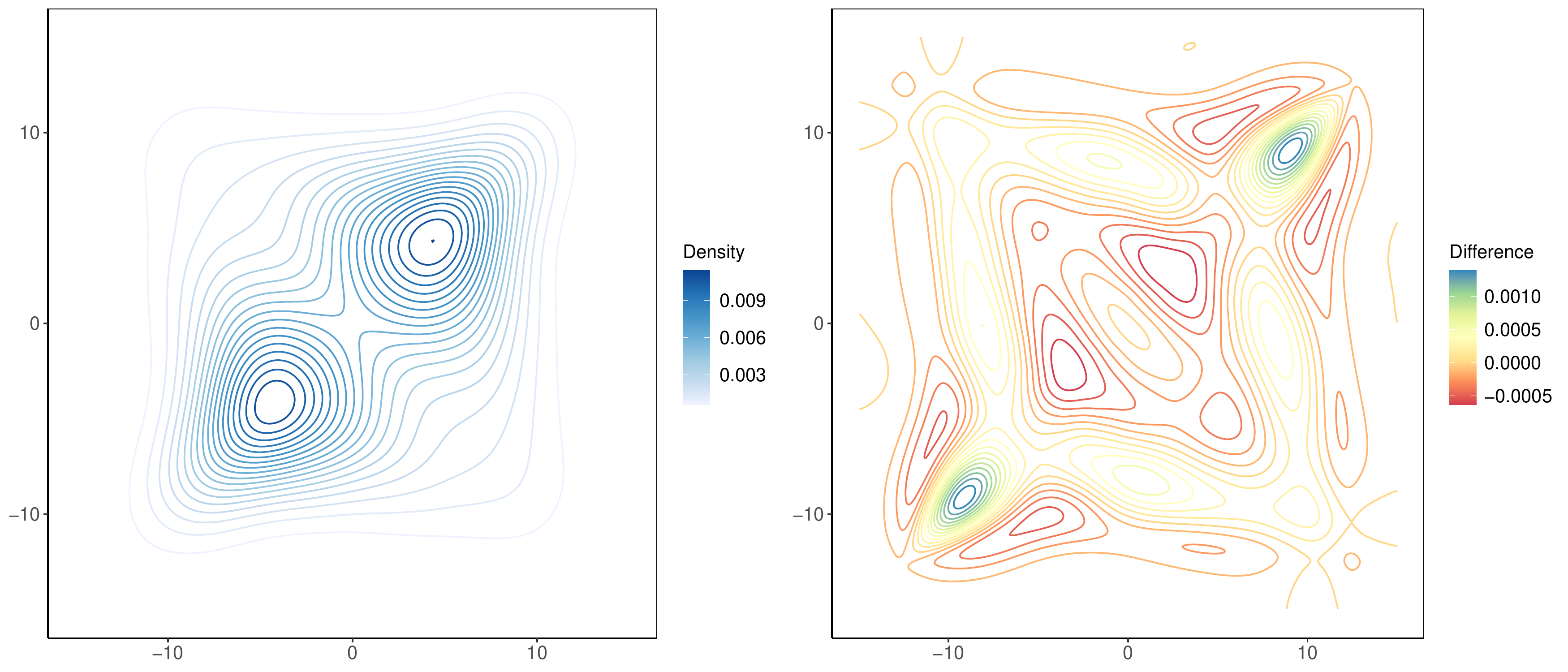}
\caption{\label{diffs} The left plot displays the estimated mixture distribution averaged over all simulation runs. The right plot shows the associated differences to the true mixture distribution.}
\end{figure}

For the nonparametric approach, we determined the optimal number of basis functions from ${n_1} = {n_2} \in \{7, \dots, 15\}$ by cross-validation. 
To ease the computational burden, we selected the same number of basis functions in both dimensions, which in the given setting is adequate as the one-dimensional marginal distributions are identical.
We used ten cross-validation partitions per simulation run and the out-of-sample likelihood of the test set (10\% of the data) to assess the fit. In more than two thirds of all simulation runs, this cross-validation led to the choice of $n_1 = n_2 \in \{9, 10, 11\}$.
The mean estimates of the diagonal entries of the t.p.m.\ were $\hat{\gamma}_{11}=0.9696$ and $\hat{\gamma}_{22}=0.9695$, respectively.
Figure~\ref{diffs} gives a visual illustration of the performance with respect to the estimation of the emission distributions, displaying the mean estimate of the overall mixture distribution resulting from the two emission distributions (averaged over all simulation runs; left panel), and additionally the corresponding differences to the true mixture distribution of the data-generating process (again the average over all runs; right panel).
As expected for our nonparametric approach, areas with negative curvature (peaks) are slightly underestimated (negative bias), while areas with positive curvature (troughs) are slightly overestimated (positive bias). The bias would decrease for increasing $n_1,n_2$, at the cost of an increased variance. 

\begin{figure}[!h]
\includegraphics[width = \textwidth]{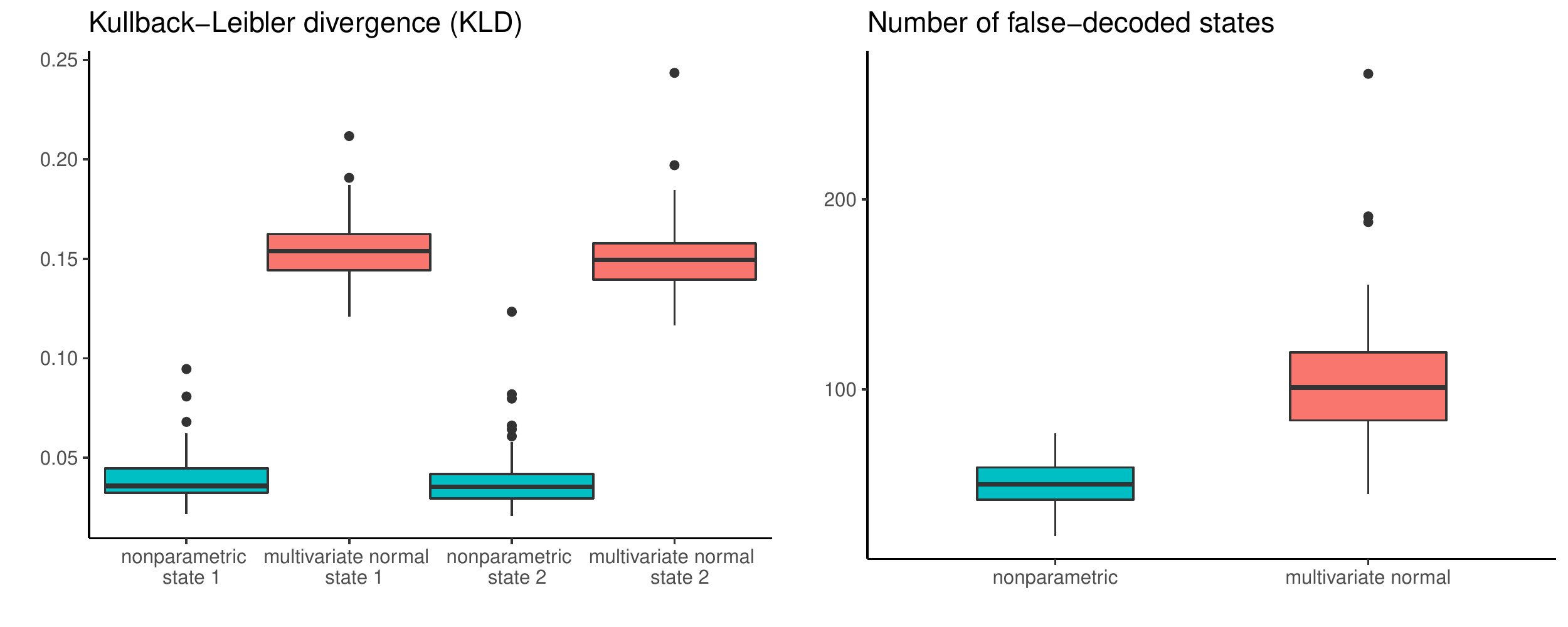}
\caption{\label{bp} On the left side, the boxplot diagrams summarise the Kullback-Leibler divergences of the nonparametric and parametric model of every simulation run for each state. On the right side, the number of false-decoded states by either approach are shown.}
\end{figure}

To compare these results to the simple parametric benchmark represented by an HMM with bivariate Gaussian emission distributions, we further calculated the Kullback-Leibler divergences (KLD) between estimated (nonparametric and parametric, respectively) and true emission distributions.  
The left panel in Figure~\ref{bp} shows boxplots of the KLDs obtained in the 100 simulation runs, which confirm the expected inferior performance of the parametric approach in the given setting. 
We additionally consider the most likely trajectory of states under the fitted models, using the Viterbi algorithm (see, e.g., \citealt{zucchini}) to find
$$ (g^{\star}_1, \ldots, g^{\star}_T) = \operatorname*{argmax}_{g_1, \ldots, g_T} \Pr (g_1, \ldots, g_T | \bold{y}_1, \ldots, \bold{y}_T).$$
The right panel in Figure~\ref{bp} gives a comparison of the state-decoding performance under the nonparametric and the parametric model, respectively. The nonparametric model delivers a proportion of $97.5$\% of correctly decoded states on average, compared to only $94.8$\% as obtained under the parametric model. 
The superior performance of the nonparametric approach is no surprise in the given setting, nevertheless it does clearly point out the potential pitfalls associated with an unfortunate choice of a parametric model specification. Put differently, in situations with complex emission distributions the nonparametric approach can substantially improve the performance, in particular with respect to state-decoding accuracy.

\subsection{Bivariate Gaussian emission distribution}

\begin{figure}[!htb]
\includegraphics[scale = 0.6]{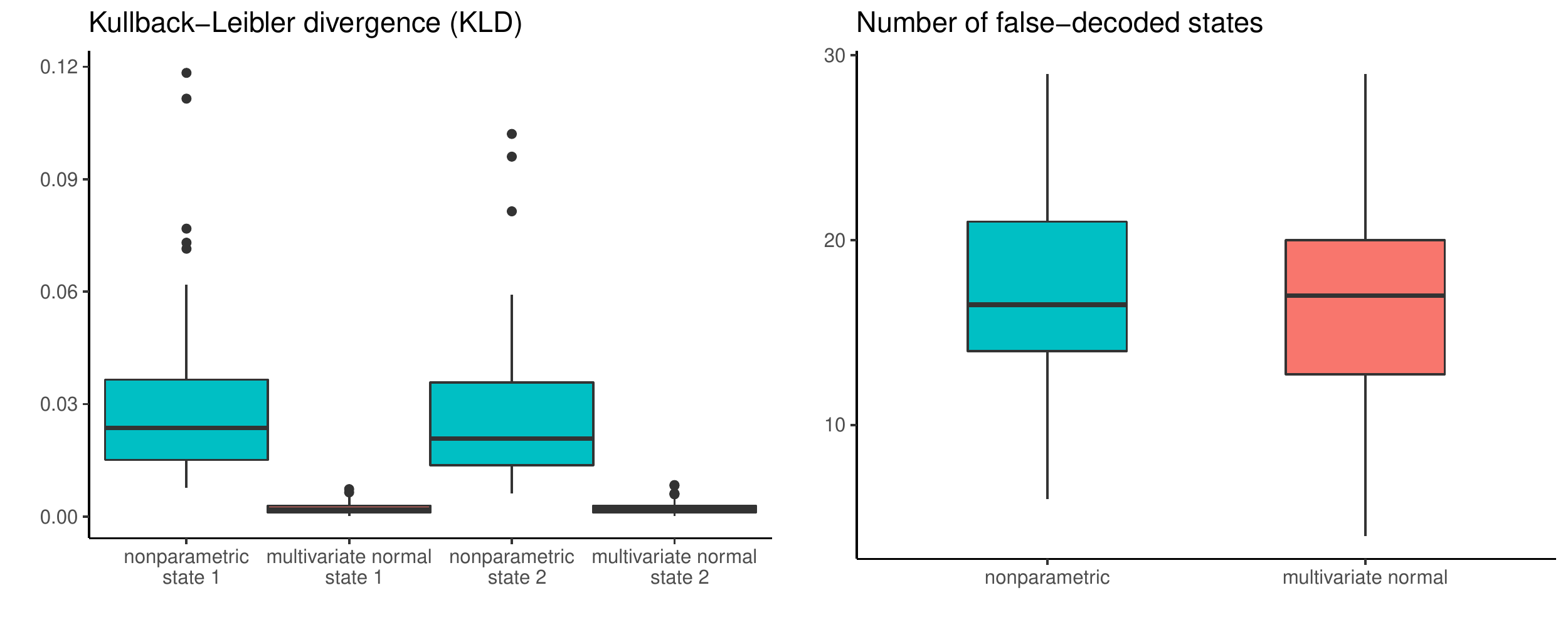}
\caption{\label{bp_norm} On the left side, the boxplot diagrams summarise the Kullback-Leibler divergences of the nonparametric and parametric model of every simulation run for each state. On the right side, the number of false-decoded states by either approach are shown.}
\end{figure}

In the second simulation experiment, we replaced the emission distributions used above by simple bivariate Gaussian distributions --- all other specifications are unchanged. The point of this exercise is to give an idea of the costs associated with using the flexible nonparametric approach when it is in fact not required. The left panel in Figure~\ref{bp_norm}, showing boxplots of the KLDs obtained under the nonparametric and the parametric approach, respectively, illustrates the (expected) superior performance of the parametric approach with respect to estimation accuracy (here of the emission distributions). The higher KLDs of the nonparametrically estimated emission distributions result from an increased variance as well as a small but systematic bias in areas of non-zero curvature.  However, remarkably, there is no notable difference in the proportion of false-decoded states between the parametric and nonparametric approach (right panel in Figure~\ref{bp_norm}). Therefore, with respect to state-decoding accuracy the cost associated with using the nonparametric approach instead of the correctly specified parametric model here is negligible. Of course it also needs to be taken into account that the computational cost is much higher when using the nonparametric approach. 

Overall, our simulations show the potential of the nonparametric approach to substantially outperform misspecified parametric models, and indicate that when compared to correctly specified parametric models the associated cost with respect to a loss of state-decoding accuracy can be negligible. Therefore, the nonparametric approach represents a potentially valuable tool for fitting HMMs to multivariate time series with complex dependence structures, or generally such where exploratory data analysis does not readily reveal an adequate candidate distribution for parametric modelling.

\section{Case study: modelling goalkeeper passes in football}\label{casestudy}

In the following case study, we demonstrate the potential practical use of the nonparametric estimation approach. Specifically, we model the lengths and angles of goalkeeper passes during the UEFA European Football Championship 2020, with the aim of linking the HMM states to a team's tactical decisions. 
A simple dichotomy of the diverse tactics to getting close to the opponent's goal (and ideally scoring a goal) is the following: either a team tries to combine their way across the field by controlled passing, or they start an attack by a long and relatively uncontrolled kick forward. Which of the two strategies is currently predominantly employed by a team can be inferred from the actions on the pitch. To this end, an HMM can be used to link observable metrics such as pass lengths to underlying tactics (see, e.g., \citealp{otting2021copula, otting2022football}). In general, studies trying to infer and interpret different tactics have seen a steady rise in recent years, driven by an increasing amount of event- and tracking data in football becoming available (see, e.g., \citealp{decroos2018automatic, robberechts2019valuing, decroos2019player,decroos2020soccermix,bauer2021data}). 

\subsection{Data}

We consider an event data set which was made publicly available by the company StatsBomb, one of the largest data providers of football data. The data set is accessible via the `StatsBombR' package (see \citealp{statsbomb}). It comprises all players' actions (e.g.\ passes, shots, tackles) during the UEFA European Football Championship 2020 (played in 2021). Goalkeepers are often the first to initiate an organised attacking sequence; for a simple illustration of our methodology, we thus make the simplifying assumption that the attacking tactics of a team can be inferred from the goalkeeper's actions. We focus on passes played by each team's goalkeeper, specifically the length and angle of those passes (comprising in-play passes, free kicks, and goal kicks). Overall, $N = 3353$ goalkeeper passes were played in the $51$ matches of the tournament. As each match involves two teams, the data set is given by $M = 102$ different time series. 
The most passes in one match $(58)$ were played by Switzerland's Yann Sommer in the quarter-final against Spain, the least $(11)$ by Kasper Schmeichel in the group-stage match between Denmark and Finland. 

\begin{figure}[!h]
  \centering
  \includegraphics[scale = 0.6]{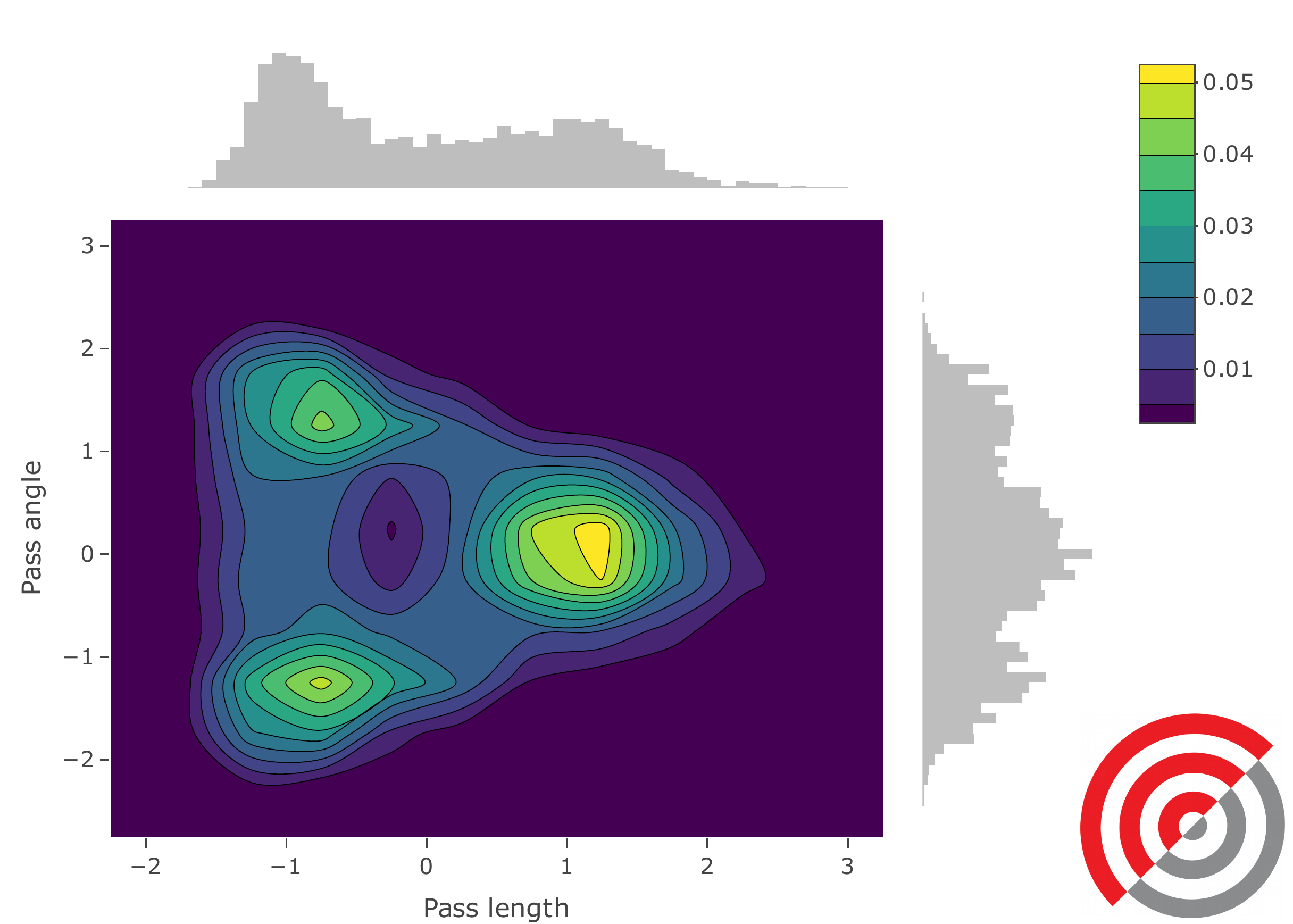}
\caption{\label{hist} The contour plot shows the joint empirical distribution of the standardised lengths and angles of the goalkeeper passes. The  histogram to the top displays the empirical distribution of the standardised pass lengths and the histogram to the right displays the empirical distribution of the standardised pass angles.
}
\end{figure}

The histograms of both empirical one-dimensional marginal distributions and a contour plot of the empirical joint distribution of pass length and pass angle are displayed in Figure~\ref{hist}. Both pass length and pass angle were standardised to have zero mean and unit variance. For the pass length, we observe a bimodal structure with passes either short (likely controlled passes to a defender) or long (predominantly relatively uncontrolled long kicks). For the pass angle, we find three modes in the empirical distribution, one for either side (left or right) and one for the centre of the pitch. Considering the fairly complex structure of the bivariate empirical distribution of pass lengths and pass angles, it is clearly difficult to conceive a suitable parametric HMM formulation. Arguably, a 3-state Gaussian HMM might be adequate in terms of the goodness-of-fit, however such a model would differentiate between short passes to left and right defenders, respectively, which is not desirable.

\subsection{Model formulation and inference}

In view of our aim to distinguish the two dominant strategies for initiating an attack --- short passes or long kicks --- we fitted a 2-state HMM to the bivariate times series of pass lengths and pass angles, estimating the bivariate emission distributions nonparametrically using tensor-product B-splines as described above. 
We numerically maximise the joint log-likelihood of the $M=102$ time series, which assuming independence across matches and goalkeepers is given by
\begin{equation*} 
\log \mathcal{L} = \sum_{m = 1}^{102} \log \bigl( \delta \mathbf{P}_1^{(m)} \boldsymbol{\Gamma} \mathbf{P}_2^{(m)} \cdot \ldots \cdot \boldsymbol{\Gamma} \mathbf{P}_T^{(m)} \mathbf{1}' \bigr).
\label{finallike}
\end{equation*}
To calculate the logarithm of the matrix product, a scaling strategy as described in \citet{zucchini} is used. The number of basis functions is again specified to be identical for the two dimensions, and using 5-fold cross-validation --- holding out ${\sim}10\%$ of the time series in each fold for testing --- was selected as $n_1=n_2=9$.

\subsection{Results}

\begin{figure}[!h]
\centering
\includegraphics[scale = 0.55]{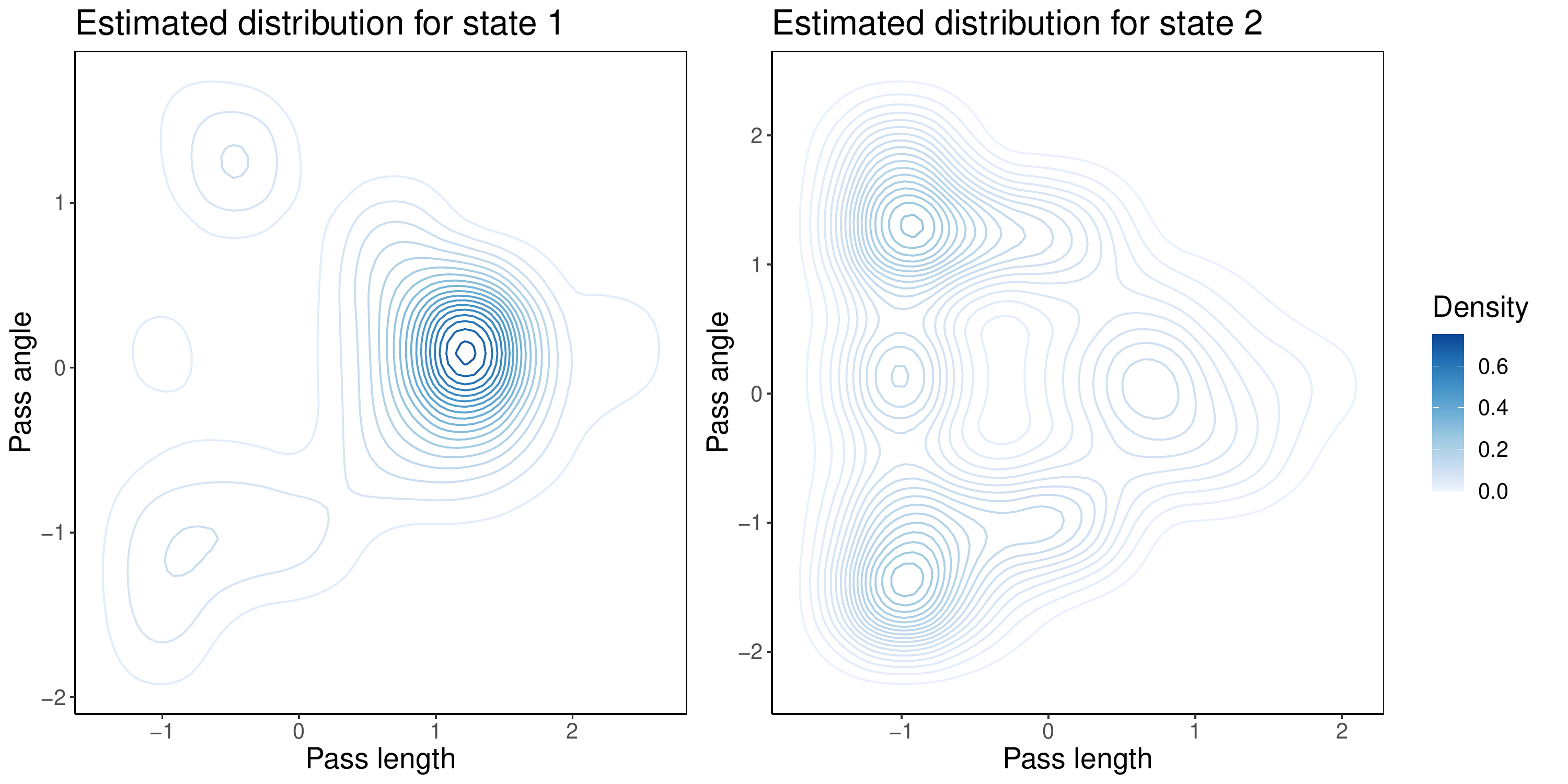}
\caption{\label{states} Contour plots showing the estimated emission distributions of the 2-state nonparametric HMM. }
\end{figure}

The estimated nonparametric emission distributions of the 2-state HMM are displayed in Figure~\ref{states}. State 1 is associated with predominantly long kicks to the centre of the pitch, whereas state 2 implies shorter passes, to defenders on both sides as well as to defensive midfielders. Both states allow for occasional deviations from the overall tactics associated with the state. 
While the states do indicate the expected main tactics for attacks, it needs to be kept in mind that the states will merely be proxies for the actual tactics due to the unsupervised training of the model.
The t.p.m.\ of the underlying Markov chain was estimated as 
\begin{equation*}
    \widehat{\Gamma} = 
\begin{pmatrix}
0.948 & 0.052\\
0.034 & 0.966
\end{pmatrix},
\end{equation*}
i.e.\ we find high persistence in the states. This seems plausible as match tactics would be expected to change only a few times within match, if at all.

\begin{figure}[!h]
\centering
\includegraphics[scale = 0.8]{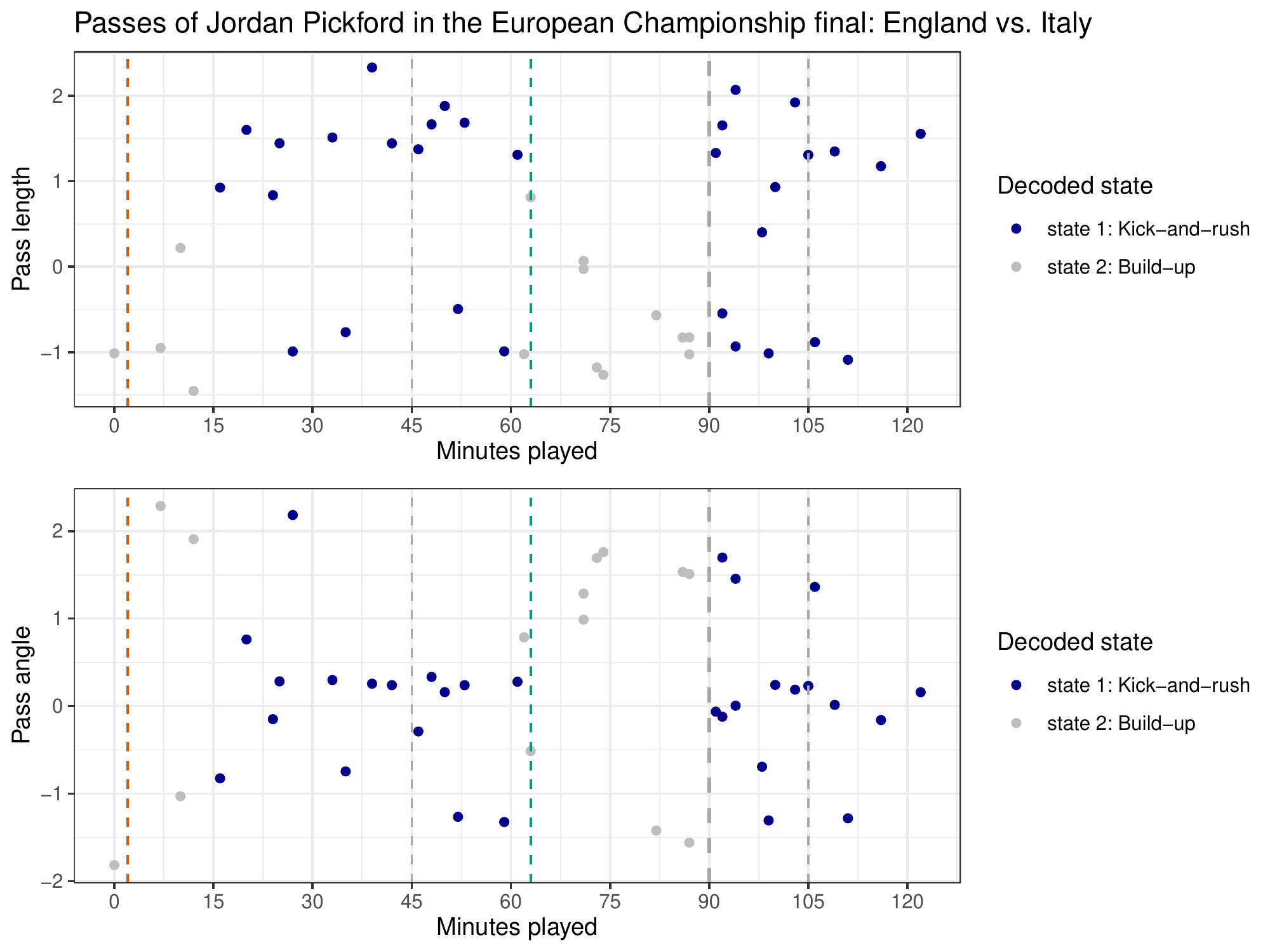}
\caption{\label{passes} 
Time series of standardised pass lengths (top) and pass angles (bottom), colour-coded according to the Viterbi-decoded states. The vertical dotted lines indicate the goals scored by England (red) and Italy (green), respectively, as well as half time and full time (grey).
}
\end{figure}

To further illustrate and interpret the fitted model, we used the Viterbi algorithm to decode the states underlying the bivariate observations. In Figure~\ref{passes}, we display the lengths and angles of the passes mady by England's goalkeeper, Jordan Pickford, during the final against Italy, colour-coded according to the Viterbi-decoded states. According to the model and the decoded states, England set out at the beginning of the match building attacks by mostly controlled passing. Likely as a consequence of their early lead, this changed about 15 minutes into the match, with England now mostly resorting to long goal kicks (kick-and-rush), presumably to reduce the risk of losing the ball close to their own goal. After Italy's equaliser in the 67th minute, England returned to more controlled build-up play, trying to regain control of the match, before again resorting to long goal kicks (and thereby lowering the risk) during the extra time. Despite obvious limitations of the relatively simplistic model applied (e.g.\ states only being proxies of actual tactics, two-state dichotomy of what likely is a more complex data-generating process), this decoded state sequence is in accordance with media reports of the match. 

\subsection{Inclusion of covariates of interest}

\begin{figure}[!h]
\centering
\includegraphics[scale = 0.9]{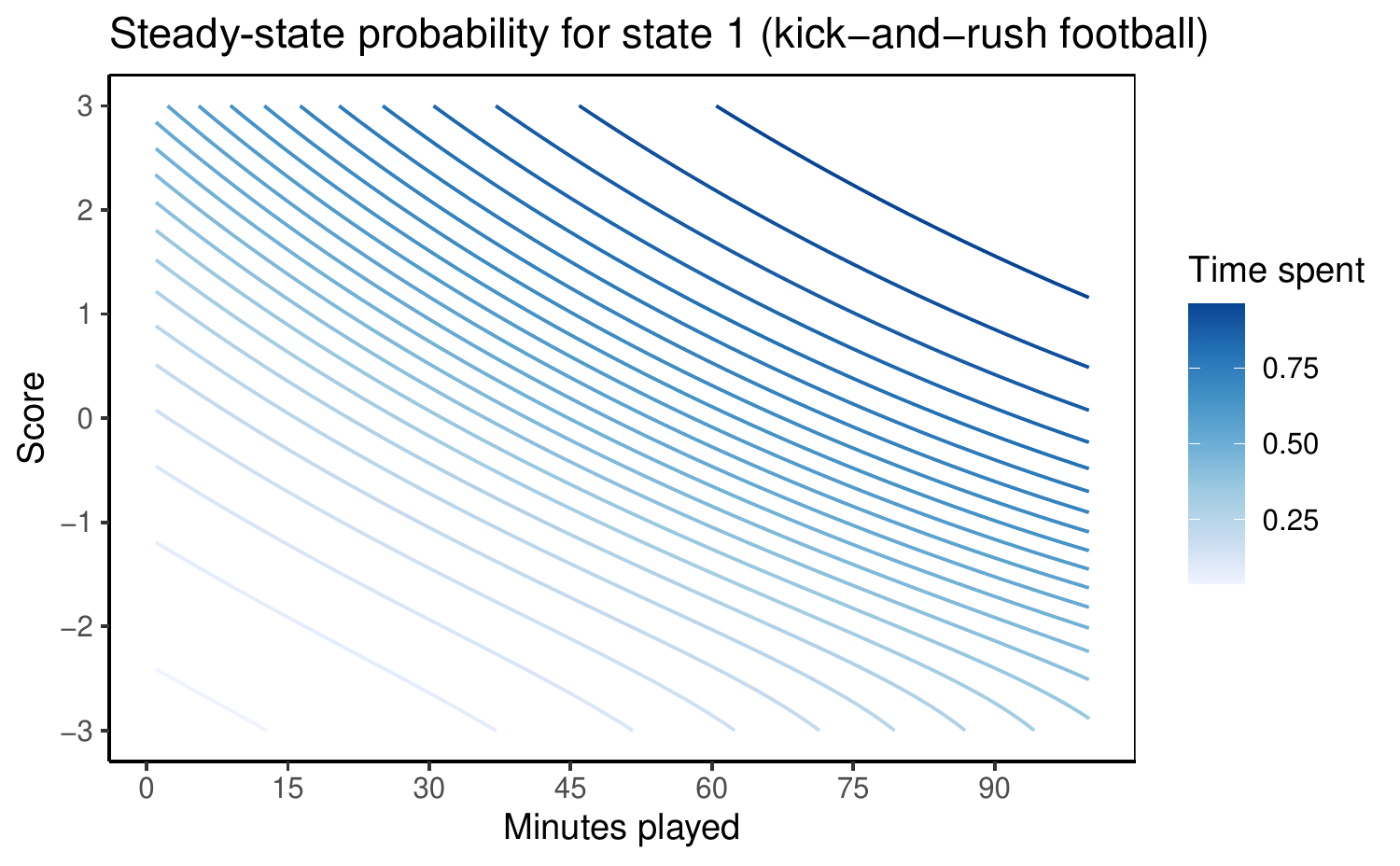}
\caption{\label{trans} The plot displays the steady-state probability
of England being in state 1 as a function of match time and the current score.}
\end{figure}

To illustrate the potential of the approach in particular for sports analytics, we consider the inclusion of covariates as an extension of the model presented above. Specifically, we include covariates in the state transition probabilities, thereby trying to account for changes in the match tactics driven by match dynamics. The now time-varying transition probabilities $\gamma_{ij}^{(t)}$, $i, j = 1, \ldots, N$, are modelled depending on the covariate vector $\bold{x}^{(t)} =  (x_{1}^{(t)}, \ldots, x_{P}^{(t)})$ using the multinomial logit link,
\begin{equation}\label{multi}
\gamma_{ij}^{(t)} = \frac{\exp(\nu_{ij}^{(t)})}{\sum_{k = 1}^{N}\exp(\nu_{ik}^{(t)})},
\end{equation} with
\begin{equation*}
\nu_{ij}^{(t)} = 
\begin{cases}
{\omega}_0^{(ij)} + \sum_{l = 1}^{P} {\omega}_l^{(ij)} x_l^{(t)} & \text{ if } i \neq j, \\
0 & \text{ otherwise}.
\end{cases}
\end{equation*}
As example covariates, we consider the current match time, the current score, and their interaction, i.e. 
$$\bold{x}^{(t)} = (minute^{(t)}, score^{(t)}, minute^{(t)} \cdot score^{(t)}).$$
The two emission distributions are again estimated nonparametrically. Figure \ref{trans} displays the steady-state probability of state 1 as a function of the match time and the current score (these are obtained from the stationary distribution the Markov chain would converge to when fixing the entries of the t.p.m.\ at the respectice combination of match time and current score; cf.\ \citealp{patterson2009classifying}). 
Teams that are currently trailing are more likely to play constructive build-up football, whereas teams in the lead tend to resort to lower-risk long goal kicks. Towards the end of the match, goalkeepers in any case are more likely to play longer balls, which for teams that are trailing can be explained by their time running out. 

This very simple example analysis showcases how inference (and predictions) on tactical decisions can be made without time-consuming video analysis. In actual sports analytics practice, more complex HMMs, in particular such that include more covariates related to the match dynamics, would typically be built in order to generate insights into an opponent's tactical considerations.

\section{Discussion}

This contribution explores a nonparametric approach for modelling the emission distributions within multivariate HMMs without the need to select a parametric family. The simulation experiments and the case study illustrate potential settings in which the approach can be useful, namely such with complex shapes of the emission distributions, potentially exacerbated by an overlap of the different states' distributions, which renders it particularly difficult to decide on a parametric family based on exploratory data analysis. Realistically, the methodology as it stands is applicable only to low-dimensional multivariate time series ($D=2,3$) due to the curse of dimensionality. In general, due to its increased complexity compared to simple parametric modelling the approach will be advantageous only in situations when no good parametric model formulation is evident. The nonparametric approach can in any case serve as a tool for exploratory data analysis, as it can be used to fit a multivariate HMM without distributional assumption, the results of which may indicate potentially adequate parametric families. Promising areas of application of the proposed approach include ecology (modelling multiple behavioural measures of an animal, see \citealp{deruiter2017multivariate}), finance (modelling concentration risks in portfolios with correlated stocks, see \citealp{maruotti2019hidden}), and medicine (modelling multiple signals from electronic health records, see \citealp{alaa2018hidden}).

Regarding the methodology, a challenge not addressed in the present contribution is the selection of the smoothing parameter(s) when estimating the nonparametric HMM using optimisation of a likelihood penalised for wiggliness (i.e.\ P-spline modelling as in \citealp{splines}). This can for example be achieved using information criteria, estimating the effective degrees of freedom based on the Fisher information \citep{langrock2018spline}. Instead of constructing the multivariate emission distributions using B-splines, alternative smoothing approaches like kernel density methods (see, e.g., \citealp{piccardi2007hidden, DEGOOIJER2022107431}) could also be used. 
Model extensions concerning the state process --- e.g.\ continuous-time formulations \citep{jackson2003multistate}, semi-Markovian processes \citep{langrock2011hidden}, and coupled state processes \citep{pohle2021primer} --- are straightforward and do not lead to additional technical challenges. 

\bibnote{statsbomb}{[dataset]}

\bibliographystyle{apalike}

\begin{thebibliography}{}

\bibitem[Ailliot et~al., 2009]{ailliot2009space}
Ailliot, P., Thompson, C., and Thomson, P. (2009).
\newblock Space--time modelling of precipitation by using a hidden {M}arkov
  model and censored gaussian distributions.
\newblock {\em Journal of the Royal Statistical Society: Series C (Applied
  Statistics)}, 58(3):405--426.

\bibitem[Alaa and Van Der~Schaar, 2018]{alaa2018hidden}
Alaa, A.~M. and Van Der~Schaar, M. (2018).
\newblock A hidden absorbing semi-{M}arkov model for informatively censored
  temporal data: learning and inference.
\newblock {\em The Journal of Machine Learning Research}, 19(1):108--169.

\bibitem[Altman, 2007]{altman2007mixed}
Altman, R. (2007).
\newblock Mixed hidden {M}arkov models: an extension of the hidden {M}arkov
  model to the longitudinal data setting.
\newblock {\em Journal of the American Statistical Association},
  102(477):201--210.

\bibitem[Bauer and Anzer, 2021]{bauer2021data}
Bauer, P. and Anzer, G. (2021).
\newblock Data-driven detection of counterpressing in professional football.
\newblock {\em Data Mining and Knowledge Discovery}, 35(5):2009--2049.

\bibitem[Beumer et~al., 2020]{beumer2020application}
Beumer, L.~T., Pohle, J., Schmidt, N.~M., Chimienti, M., Desforges, J.-P.,
  Hansen, L.~H., Langrock, R., Pedersen, S.~H., Stelvig, M., and van Beest,
  F.~M. (2020).
\newblock An application of upscaled optimal foraging theory using hidden
  {M}arkov modelling: year-round behavioural variation in a large arctic
  herbivore.
\newblock {\em Movement Ecology}, 8(1):1--16.

\bibitem[Brunel and Pieczynski, 2005]{brunel2005unsupervised}
Brunel, N. and Pieczynski, W. (2005).
\newblock Unsupervised signal restoration using hidden {M}arkov chains with
  copulas.
\newblock {\em Signal Processing}, 85(12):2304--2315.

\bibitem[Bulla et~al., 2012]{bulla2012multivariate}
Bulla, J., Lagona, F., Maruotti, A., and Picone, M. (2012).
\newblock A multivariate hidden {M}arkov model for the identification of sea
  regimes from incomplete skewed and circular time series.
\newblock {\em Journal of Agricultural, Biological, and Environmental
  Statistics}, 17(4):544--567.

\bibitem[Celeux and Durand, 2008]{celeux2008selecting}
Celeux, G. and Durand, J.-B. (2008).
\newblock Selecting hidden {M}arkov model state number with cross-validated
  likelihood.
\newblock {\em Computational Statistics}, 23(4):541--564.

\bibitem[{De Gooijer} et~al., 2022]{DEGOOIJER2022107431}
{De Gooijer}, J.~G., Henter, G.~E., and Yuan, A. (2022).
\newblock Kernel-based hidden {M}arkov conditional densities.
\newblock {\em Computational Statistics \& Data Analysis}, 169:107431.

\bibitem[Decroos and Davis, 2019]{decroos2019player}
Decroos, T. and Davis, J. (2019).
\newblock Player vectors: Characterizing soccer players’ playing style from
  match event streams.
\newblock In {\em Joint European {C}onference on {M}achine {L}earning and
  {K}nowledge {D}iscovery in {D}atabases}, pages 569--584. Springer.

\bibitem[Decroos et~al., 2018]{decroos2018automatic}
Decroos, T., Van~Haaren, J., and Davis, J. (2018).
\newblock Automatic discovery of tactics in spatio-temporal soccer match data.
\newblock In {\em Proceedings of the 24th ACM SIGKDD International Conference
  on Knowledge Discovery \& Data Mining}, pages 223--232.

\bibitem[Decroos et~al., 2020]{decroos2020soccermix}
Decroos, T., van Roy, M., and Davis, J. (2020).
\newblock Soccermix: Representing soccer actions with mixture models.
\newblock In {\em Joint European {C}onference on {M}achine {L}earning and
  {K}nowledge {D}iscovery in {D}atabases}, pages 459--474. Springer.

\bibitem[DeRuiter et~al., 2017]{deruiter2017multivariate}
DeRuiter, S., Langrock, R., Skirbutas, T., Goldbogen, J., Calambokidis, J.,
  Friedlaender, A., and Southall, B.~L. (2017).
\newblock A multivariate mixed hidden {M}arkov model for blue whale behaviour
  and responses to sound exposure.
\newblock {\em The Annals of Applied Statistics}, 11(1):362--392.

\bibitem[Eilers et~al., 2006]{eilers2006fast}
Eilers, P., Currie, I., and Durb{\'a}n, M. (2006).
\newblock Fast and compact smoothing on large multidimensional grids.
\newblock {\em Computational Statistics \& Data Analysis}, 50(1):61--76.

\bibitem[Eilers and Marx, 1996]{splines}
Eilers, P. H.~C. and Marx, B.~D. (1996).
\newblock {Flexible smoothing with B-splines and penalties}.
\newblock {\em Statistical Science}, 11(2):89 -- 121.

\bibitem[Fahrmeir et~al., 2013]{epub31413}
Fahrmeir, L., Kneib, T., Lang, S., and Marx, B. (2013).
\newblock {\em Regression: Models, Methods and Applications}.
\newblock Springer-Verlag, Berlin.

\bibitem[Fiecas et~al., 2017]{fiecas2017shrinkage}
Fiecas, M., Franke, J., von Sachs, R., and Tadjuidje~Kamgaing, J. (2017).
\newblock Shrinkage estimation for multivariate hidden {M}arkov models.
\newblock {\em Journal of the American Statistical Association},
  112(517):424--435.

\bibitem[H{\"a}rdle et~al., 2015]{hardle2015hidden}
H{\"a}rdle, W.~K., Okhrin, O., and Wang, W. (2015).
\newblock Hidden {M}arkov structures for dynamic copulae.
\newblock {\em Econometric Theory}, 31(5):981--1015.

\bibitem[Jackson et~al., 2003]{jackson2003multistate}
Jackson, C.~H., Sharples, L.~D., Thompson, S.~G., Duffy, S.~W., and Couto, E.
  (2003).
\newblock Multistate markov models for disease progression with classification
  error.
\newblock {\em Journal of the Royal Statistical Society: Series D (The
  Statistician)}, 52(2):193--209.

\bibitem[Lanchantin et~al., 2011]{lanchantin2011unsupervised}
Lanchantin, P., Lapuyade-Lahorgue, J., and Pieczynski, W. (2011).
\newblock Unsupervised segmentation of randomly switching data hidden with
  non-gaussian correlated noise.
\newblock {\em Signal Processing}, 91(2):163--175.

\bibitem[Langrock et~al., 2018]{langrock2018spline}
Langrock, R., Adam, T., Leos-Barajas, V., Mews, S., Miller, D.~L., and
  Papastamatiou, Y.~P. (2018).
\newblock Spline-based nonparametric inference in general state-switching
  models.
\newblock {\em Statistica Neerlandica}, 72(3):179--200.

\bibitem[Langrock et~al., 2015]{langrock2015nonparametric}
Langrock, R., Kneib, T., Sohn, A., and DeRuiter, S. (2015).
\newblock Nonparametric inference in hidden {M}arkov models using {P}-splines.
\newblock {\em Biometrics}, 71(2):520--528.

\bibitem[Langrock and Zucchini, 2011]{langrock2011hidden}
Langrock, R. and Zucchini, W. (2011).
\newblock Hidden markov models with arbitrary state dwell-time distributions.
\newblock {\em Computational Statistics \& Data Analysis}, 55(1):715--724.

\bibitem[Maruotti et~al., 2019]{maruotti2019hidden}
Maruotti, A., Punzo, A., and Bagnato, L. (2019).
\newblock Hidden {M}arkov and semi-{M}arkov models with multivariate
  leptokurtic-normal components for robust modeling of daily returns series.
\newblock {\em Journal of Financial Econometrics}, 17(1):91--117.

\bibitem[Ng{\^o} et~al., 2019]{ngo2019understanding}
Ng{\^o}, M.~C., Heide-J{\o}rgensen, M.~P., and Ditlevsen, S. (2019).
\newblock Understanding narwhal diving behaviour using hidden {M}arkov models
  with dependent state distributions and long range dependence.
\newblock {\em PLoS Computational Biology}, 15(3):e1006425.

\bibitem[Orfanogiannaki and Karlis, 2018]{orfanogiannaki2018multivariate}
Orfanogiannaki, K. and Karlis, D. (2018).
\newblock Multivariate {P}oisson hidden {M}arkov models with a case study of
  modelling seismicity.
\newblock {\em Australian \& New Zealand Journal of Statistics},
  60(3):301--322.

\bibitem[{\"O}tting and Karlis, 2022]{otting2022football}
{\"O}tting, M. and Karlis, D. (2022).
\newblock Football tracking data: a copula-based hidden markov model for
  classification of tactics in football.
\newblock {\em Annals of Operations Research}, in press.

\bibitem[{\"O}tting et~al., 2021]{otting2021copula}
{\"O}tting, M., Langrock, R., and Maruotti, A. (2021).
\newblock A copula-based multivariate hidden {M}arkov model for modelling
  momentum in football.
\newblock {\em AStA Advances in Statistical Analysis}, in press.

\bibitem[Patterson et~al., 2009]{patterson2009classifying}
Patterson, T.~A., Basson, M., Bravington, M.~V., and Gunn, J.~S. (2009).
\newblock Classifying movement behaviour in relation to environmental
  conditions using hidden {M}arkov models.
\newblock {\em Journal of Animal Ecology}, 78(6):1113--1123.

\bibitem[Phillips et~al., 2015]{phillips2015objective}
Phillips, J.~S., Patterson, T.~A., Leroy, B., Pilling, G.~M., and Nicol, S.~J.
  (2015).
\newblock Objective classification of latent behavioral states in bio-logging
  data using multivariate-normal hidden {M}arkov models.
\newblock {\em Ecological Applications}, 25(5):1244--1258.

\bibitem[Piccardi and P{\'e}rez, 2007]{piccardi2007hidden}
Piccardi, M. and P{\'e}rez, {\'O}. (2007).
\newblock Hidden {M}arkov models with kernel density estimation of emission
  probabilities and their use in activity recognition.
\newblock In {\em 2007 IEEE Conference on Computer Vision and Pattern
  Recognition}, pages 1--8. IEEE.

\bibitem[Pohle et~al., 2021]{pohle2021primer}
Pohle, J., Langrock, R., Schaar, M. v.~d., King, R., and Jensen, F.~H. (2021).
\newblock A primer on coupled state-switching models for multiple interacting
  time series.
\newblock {\em Statistical Modelling}, 21(3):264--285.

\bibitem[Punzo and Maruotti, 2016]{punzo2016clustering}
Punzo, A. and Maruotti, A. (2016).
\newblock Clustering multivariate longitudinal observations: The contaminated
  {G}aussian hidden {M}arkov model.
\newblock {\em Journal of Computational and Graphical Statistics},
  25(4):1097--1098.

\bibitem[Robberechts, 2019]{robberechts2019valuing}
Robberechts, P. (2019).
\newblock Valuing the art of pressing.
\newblock In {\em Proceedings of the StatsBomb Innovation In Football
  Conference}, pages 1--11. StatsBomb.

\bibitem[Ruppert et~al., 2009]{ruppert2009semiparametric}
Ruppert, D., Wand, M.~P., and Carroll, R.~J. (2009).
\newblock Semiparametric regression during 2003--2007.
\newblock {\em Electronic Journal of Statistics}, 3:1193.

\bibitem[Spezia, 2010]{spezia2010bayesian}
Spezia, L. (2010).
\newblock Bayesian analysis of multivariate gaussian hidden {M}arkov models
  with an unknown number of regimes.
\newblock {\em Journal of Time Series Analysis}, 31(1):1--11.

\bibitem[StatsBomb, 2020]{statsbomb}
StatsBomb (2020).
\newblock Stats{B}omb{R}.
\newblock R package version 0.1.0.

\bibitem[van Beest et~al., 2019]{van2019classifying}
van Beest, F.~M., Mews, S., Elkenkamp, S., Schuhmann, P., Tsolak, D., Wobbe,
  T., Bartolino, V., Bastardie, F., Dietz, R., von Dorrien, C., et~al. (2019).
\newblock Classifying grey seal behaviour in relation to environmental
  variability and commercial fishing activity-a multivariate hidden {M}arkov
  model.
\newblock {\em Scientific Reports}, 9(1):1--14.

\bibitem[Visser et~al., 2002]{visser2002fitting}
Visser, I., Raijmakers, M.~E., and Molenaar, P. (2002).
\newblock Fitting hidden {M}arkov models to psychological data.
\newblock {\em Scientific Programming}, 10(3):185--199.

\bibitem[Zimmerman et~al., 2022]{zimmerman2022copula}
Zimmerman, R., Craiu, R.~V., and Leos-Barajas, V. (2022).
\newblock Copula modelling of serially correlated multivariate data with hidden
  structures.
\newblock {\em arXiv preprint arXiv:2207.04127}, Unpublished results.

\bibitem[Zucchini et~al., 2016]{zucchini}
Zucchini, W., MacDonald, I.~L., and Langrock, R. (2016).
\newblock {\em Hidden {M}arkov {M}odels for {T}ime {S}eries: {A}n
  {I}ntroduction {U}sing {R}}.
\newblock Boca Raton: Chapman \& Hall/CRC.

\end{thebibliography}

\end{spacing}
\end{document}